\documentclass[12pt]{article}

\usepackage{latexsym}

\font\tenmsbm=msbm10 scaled 1200
\font\sevenmsbm=msbm9
\newfam\msbmfam
\textfont\msbmfam=\tenmsbm \scriptfont\msbmfam=\sevenmsbm

\makeatletter
\@addtoreset{equation}{section}
\makeatother

\newcommand{\eref}[1]{(\ref{#1})}

\def\be{\begin{equation}}
\def\ee{\end{equation}}
\def\ba{\begin{eqnarray}}
\def\ea{\end{eqnarray}}
\def\bet{\begin{tabular}}
\def\eet{\end{tabular}}

\def\nn{\nonumber}
\def\ve{\varepsilon}

\def\a{\alpha}
\def\b{\beta}
\def\g{\gamma}
\def\G{\Gamma}

\def\om{\omega}
\def\La{\Lambda}
\def\d{\delta}
\def\la{\lambda}

\def\ra{\rightarrow}

\def\ul{\underline}
\def\w{\widetilde}

\catcode`@=11

\begin{document}

\begin{titlepage}

\begin{flushright}
Preprint DFPD 03/TH/08\\
February 2008\\
\end{flushright}

\vspace{2.5truecm}

\begin{center}

{\Large \bf Superspace formulations of ten--dimensional supergravity} \vskip0.3truecm

\vspace{2.0cm}

Kurt Lechner\footnote{kurt.lechner@pd.infn.it} and Mario
Tonin\footnote{mario.tonin@pd.infn.it}

\vspace{2cm}

 {
\it Dipartimento di Fisica, Universit\`a degli Studi di Padova,

\smallskip

and

\smallskip

Istituto  Nazionale di Fisica Nucleare, Sezione di Padova,

Via F. Marzolo, 8, 35131 Padova, Italia
}

\vspace{2.5cm}

\begin{abstract}

We present a new formulation for $N=1$, $D=10$ supergravity in superspace, in presence of a Lorentz Chern--Simons--form. This formulation entails the following properties:
it furnishes a solution of the Bianchi identities that is algebraically consistent to all orders in $\a'$; at first order it is the simplest formulation proposed so far, and it
is therefore most suitable for an explicit higher order analysis; it allows a well defined perturbative expansion in $\a'$, in which no poltergeist fields appear;
it reconciles the two different classes of {\it first order} solutions available in the literature, that until now appeared physically inequivalent.

 \vspace{0.5cm}

\end{abstract}

\end{center}
\end{titlepage}

\newpage

\baselineskip 8 mm


\section{Introduction}

In the low energy limit type $I$ and heterotic strings are described by $N=1$, $D=10 $ supergravity theories, coupled to a super--Yang--Mills multiplet. From a phenomenological point of view these string theories appear appealing, since non--abelian gauge fields
appear from the beginning. On the other hand in these theories the cancelation of chiral anomalies requires to impose the modified Bianchi identity on the three--form curvature
$H$,
 \be
 \label{bian} dH=\a({\rm tr}\,FF-{\rm tr}\,RR),
 \ee
where $F$ and $R$ are respectively the Yang--Mills and Lorentz curvature two--forms, and
$\a=\a'/4$. As it stands this identity breaks supersymmetry, and the problem of its
restoration has been attacked in a series of papers in the past, \cite{Bergs}--\cite{11}; for earlier work see \cite{suym}.
The main concern of the present paper regards again the supersymmetrization of the above
Bianchi identity, in the framework of superspace that represents an efficient and
algebraically powerful approach.

The reasons for why in this paper we come back to this problem, presenting a new
superspace solution of the Bianchi--identity \eref{bian}, are the following. First of all the supergravity theory in consideration arises as the low energy effective field theory
of string theory, for which there exists now a manifestly supersymmetric and covariant
quantization scheme, the pure spinor approach \cite{berko}, and this allows in principle
to derive this effective theory perturbatively in a manifestly supersymmetric form in
superspace, \cite{berkhowe}--\cite{tonchand}. In particular in \cite{berkhowe,chand} the pure spinor approach was applied to derive pure supergravity in superspace, while in ref. \cite{tonchand}, that is based on previous work on the Green--Schwarz heterotic string sigma model \cite{mario}, it was applied to derive the Chern--Simons--induced corrections to pure supergravity, again in superspace. For an earlier pure spinor derivation of ten dimensional supergravity see \cite{howe}.
These developments, in turn, will allow for the first time to compare results from algebraic supersymmetrization techniques, with results from a
classical loop $\a'$--expansion in string theory, in a case where {\it both} are
manifestly supersymmetric, i.e. formulated in superspace. The knowledge of the
supersymmetric structure of the $\a'$--corrections to the low energy field theory,
leading in particular to corrections in powers of the Riemann tensor, play moreover a
fundamental role in flux compactifications of heterotic and type--I string theories, see
e.g. \cite{dallagata1}--\cite{blumen}. The supersymmetric completion of \eref{bian} leads in the action indeed to terms of the type $\a'R^2$, $\a'^3R^4$ etc.

The problem of the supersymmetrization of \eref{bian} in superspace has been attacked in the literature essentially through two types of methods, called in the following ``perturbative
approach'' \cite{gates1}--\cite{gates3}, and ``non--perturbative approach''
\cite{Torino}--\cite{11}.  According to the first method one tries to solve the superspace Bianchi identity \eref{bian} order by order in $\a'$, regardless of the algebraic
consistency of the entire procedure, while in the second method one derives an {\it
algebraically consistent} set of closed non linear relations for all auxiliary
superfields, that solve the Bianchi identity exactly to all orders in $\a'$; it is then a mere technical exercise to solve these relations order by order in $\a'$.
The non--perturbative approach is based on a crucial theorem, the Bonora--Pasti--Tonin (BPT)--theorem \cite{BPT}, that guarantees the algebraic consistency of the entire construction to all orders in $\a'$. The main
discrepancies between the two methods, as developed so far, are the following two: I) the rather simple {\it first order} solution furnished by the perturbative approach appears in disagreement with the one furnished by the non--perturbative approach \cite{raciti}; II) in the perturbative method a solution of \eref{bian} in the (0,4)--sector, i.e. the sector with
$0$ bosonic and $4$ fermionic vielbeins, is claimed to extend automatically to a complete solution of
\eref{bian} also in the other sectors \cite{gates2}, while the non--perturbative method reveals that this statement is true only at first order in $\a'$: at order $\a'^2$, for example, there are simple solutions in the (0,4)--sector that do not extend to solutions of the whole Bianchi--identity.

One of the goals achieved by the new superspace solution of \eref{bian} presented in the present
paper  -- in the framework of the non--perturbative approach -- is the elimination of the discrepancy I) between the two approaches. This is achieved trough a series of non trivial superfield redefinitions -- involving also a redefinition of the
two--superform potential $B$ -- that lead to a new, but physically equivalent, realization of  the BPT--theorem, and bring eventually the two different first--order solutions
to coincide. A part from closing a debate, that some time ago ran for several years, this
result establishes a well--defined and unique first order supersymmetric heterotic
effective action, on which higher order $\a'$ corrections can be firmly based. Moreover,
the first--order superspace parametrizations emerging from our exact solution appear simpler then all first--order solutions proposed so far, and they are thus particularly
suitable as starting point for the derivation of higher order corrections.

A further asset of our new solution of \eref{bian} is represented by the fact that the
equations of motion of the physical fields do not propagate poltergeist degrees of freedom, and that the equations for the auxiliary fields admit a unique perturbative solution, as expansions in $\a'$.

The results of this paper confirm, on the other hand, the discrepancy II) mentioned above: our new solution confirms the point of the non--perturbative approach, as we will show explicitly in the case of the $\a'^2$--corrections to the $H$--constraints.

Recently there has been a proposal for the inclusion of $\a'^2$--corrections \cite{reilly}, that relaxes the ``classical'' torsion constraint,
 \be\label{class}
T_{\a\b}{}^a=2\G^a_{\a\b},
 \ee
allowing for the presence of a 1050 irreducible representation of $SO(1,9)$ in $T_{\a\b}{}^a$ \cite{nils1050}. In this  case the whole framework would change. Our viewpoint in the present paper is conservative in that we maintain the classical constraint, that is also kept in the perturbative approach. We will comment briefly on the $\a'^2$--corrections proposed in \cite{reilly} in the concluding section.
For the sake of clarity we stress that the superspace framework presented in this paper -- based on \eref{class} -- can clearly
not accommodate the entire string effective action. For example, when one takes string loop--corrections into account, like the terms that cancel the Green--Schwarz anomaly, then the r.h.s. of \eref{class} must acquire necessarily a 1050 irrep. of order $\a'^3$ \cite{candlech}. In this case the Bianchi--identity \eref{h7} is indeed no longer valid, being replaced by $dH_7=\a'^3X_8$ where $X_8$ is the standard anomaly polynomial.

Our solution is based on a set of kinematical superspace constraints that is characterized by the fact that, at zero order in $\a'$ the Yang--Mills curvature
$F$ and the supergravity curvature $R_a{}^b$ are parametrized formally in an identical manner, obeying both the constraints,
 \be\label{fr}
F_{\a\b}=0=R_{\a\b a}{}^b.
 \ee
This choice is particularly useful in that it allows on one hand a close comparison with the component level results \cite{Bergs}, that are based heavily on this tight analogy between the two sectors, and on the other hand this choice is a suitable zero--order starting point for the derivation of the superspace effective action in the framework of the pure--spinor approach, proposed recently in \cite{chand}--\cite{tonchand}.

The paper is organized as follows. In the next section we present our choice of    kinematical torsion constraints and illustrate its compatibility with the existence of
a three--form $H$ and its dual seven--form $H_7$. In section three we recall the   BPT--theorem and present its new realization. Section four is devoted to a comparison of the -- apparently contradictory -- first order results of \cite{gates2} and \cite{raciti}, on the basis of our new solution. Actually, the set of kinematical constraints used in the present paper differs from the ones of \cite{raciti} by a simple shift of the vectorial connection. Section five is devoted to concluding remarks.

\section{A set of kinematical constraints}

The starting point in a superspace approach to supergravity is the choice of a set of           kinematical constraints on the supertorsion two--form,
$$
T^A= DE^A=dE^A + E^B\Omega_B{}^A,
$$
where the one--forms $E^A=dZ^M E_M{}^A(Z)=(E^a,E^\a)$ indicate respectively the bosonic and fermionic super--zehnbeins, with $a=(0,\cdots,9)$ and $\a=(1,\cdots,16)$. The basic identity to solve is the torsion identity,
 \be\label{tbian}
DT^A=E^BR_B{}^A,
 \ee
where $R_A{}^B=d\Omega_A{}^B+\Omega_A{}^C\Omega_C{}^B$ indicates the supercurvature two--form, whereas the curvature identity $DR_A{}^B=0$ follows from \eref{tbian}, see \cite{dragon}. Recall that we have $R_a{}^\a=0=R_\a{}^a$ and,
$$
R_\a{}^\b={1\over 4}\,(\G_{ab})_\a{}^\b R^{ab}.
$$
Imposing solely the rigid torsion constraint \eref{class} and using the techniques developed in \cite{constr,candiello}, it can be shown that a set of kinematical constraints solving \eref{tbian} is given by,
 \ba
T_{\a\b}{}^a&=&2\G^a_{\a\b},\label{1}\\
T_{\a a}{}^b&=& 0,\\
T_{\a\b}{}^\g&=& 2\d^\g_{(\a}\la_{\b)}-\G^a_{\a\b}(\G_a)^{\g\d}\la_\d,\\
T_{a\a}{}^\b&=&{1\over 4}\left((\G^{bc})_\a{}^\b T_{abc}+(\G_a)_{\a\d}S^{\d\b}\right),\\
R_{\a\b ab}&=&(\G_{[a})_{\a\g}S^{\g\d}(\G_{b]})_{\d\b},\label{5}\\
R_{a\a bc}&=&2(\G_a)_{\a\b} T_{bc}{}^\b-{3\over 2}\,(\G_{[a})_{\a\b}S_{bc]}{}^\b,\label{6}
\ea
where we have the spinorial derivatives,
 \ba
D_\a\la_\b&=&-\G^a_{\a\b}D_a \phi+\la_\a\la_\b+ {1\over12}(\G_{abc})_{\a\b}\left(T^{abc}-6S^{abc}\right),\\
D_\a T_{abc}&=& (\G_{[a})_{\a\b}\left(-6T_{bc]}{}^\b+3S_{bc]}{}^\b\right),\\
D_\a S^{\b\g}&=&-2T_{\a\d}{}^{[\b} S^{\g]\d}+(\G^{ab})_\a{}^{[\b}S_{ab}{}^{\g]}.\label{9}
 \ea
In the Yang--Mills sector we have the Bianchi--identity,
 \be\label{ym}
 DF=0,
 \ee
with the standard solution,
 \ba
F_{\a\b}&=&0,\label{ym1}\\
F_{a\a}&=&2(\G_a)_{\a\b}\chi^\b,\label{ym2}\\
D_\a \chi^\b&=&T_{\a\ve}{}^\b \chi^\ve+{1\over 4}\,(\G^{cd})_\a{}^\b F_{cd},\label{ym3}
 \ea
where $\chi^\a$ is the gluino superfield. As usual $\phi$ indicates the dilaton and $\la_\a=D_\a\phi$ the gravitello superfield.

As any conventional set of constraints in $N=1$, $D=10$ supergravity, the parametrizations \eref{1}--\eref{9} are characterized by two antisymmetric third rank auxiliary tensors, $T^{abc}$ and $S^{abc}$, where we defined also the antisymmetric bi--spinor,
$$
S^{\a\b}=(\G_{abc})^{\a\b}S^{abc}\quad \leftrightarrow \quad S^{abc}={1\over 96} \, (\G^{abc})_{\a\b}S^{\a\b}.
$$
As we will see, in the present framework these two fields play the following roles:
the vectorial torsion $T^{abc}$ identifies the purely bosonic components of the three--form field--strength, see \eref{ht}, while $S^{\a\b}$ plays the role of an external "current", i.e. of a composed field that describes the (self)interactions of the supergravity multiplet.

In the formulae above a part from the fields already mentioned there appear two more fields, one is $T_{ab}{}^\a$ that is the field--strength of the gravitino, and the other is $S_{ab}{}^\a$ that is, however, completely fixed by the spinorial derivative of $S^{\a\b}$, see \eref{9}. This implies that the formulae \eref{1}--\eref{9} give a complete dynamical description of the supergravity theory, once the auxiliary superfield $S^{\a\b}$ is specified in terms of the physical fields. In particular, the choice,
$$
S^{\a\b}=0=S_{ab}{}^\a,
$$
amounts to pure $N=1$, $D=10$ supergravity. Notice also that our choice of kinematical constraints has the particular virtue, that in pure supergravity the gravitational curvature two--form $R_a{}^b$ is parameterized formally in exactly the same way as the Yang--Mills curvature $F$ -- see \eref{5}, \eref{6} and  \eref{ym1}, \eref{ym2} -- via the identification $\chi^\a \leftrightarrow T_{ab}{}^\a$.

As we will now recall, there is no need to impose additional constraints on the three--form superfield -- or its dual $H_7$ -- as the form of these fields as well as the equations of motion for all physical fields, follow already from \eref{1}--\eref{9}. To clarify this point we introduce first a class of four--superforms that will play a crucial role in what follows.

\subsection{Bianchi identities for the 3--form and 7--seven form fields}

\ul{\it A particular class of four--superforms.} In general an arbitrary $p$--superform $W_p$ can be decomposed in sectors $(m,n)$, according to the number of bosonic $(E^a)$ and fermionic $(E^\a)$ super--zehnbeins. We will write this decomposition as,
$$
W_p={1\over p!}\,E^{A_1}\cdots E^{A_p} W_{A_p\cdots A_1}=\sum _{m+n=p} W_{m,n}.
$$
For a three--form we write, for example, $W=W_{0,3}+W_{1,2}+W_{2,1}+W_{3,0}$, where,
$$
W_{1,2}={1\over 2}\,E^a E^\a E^\b W_{\b\a a}, \quad etc.
$$
We introduce then the space $V_4$ of closed four--forms $W$, defined by,
 \be\label{v4}
V_4\equiv \left\{W:\, dW=0, \quad W_{0,4}=0=W_{1,3}, \quad W_2= {1\over 2}\,E^bE^aE^\a E^\b (\G_{[a})_{\a\g}L^{\g\d}(\G_{b]})_{\d\b}\right\},
 \ee
where $L^{\a\b}$ is an antisymmetric bi--spinor.
It can be shown that the components $W_{3,1}$ and $W_{4,0}$ are uniquely determined by $W_{2,2}$, i.e. by $L^{\a\b}$, thanks to the constraint $dW=0$, see \cite{BPT,puzzle}. A form $W$ belonging to $V_4$ is thus {\it completely specified} given the tensor $L^{\a\b}$.

Following \cite{candiello} one can now ``reconstruct" the three-- and seven--form curvatures of ten--dimensional supergravity. We {\it define} a three--form $\w H$ with components,
 \ba
\w H_{\a\b\g}&=&0,\label{h03}\\
\w H_{\a\b a}&=& 2(\G_a)_{\a\b},\\
\w H_{\a a b}&=&0,\\
\w H_{abc}&=&T_{abc},\label{ht}
 \ea
and a seven--form $H_7$ with components,
 \ba
 H_{0,7}&=&\cdots =H_{4,3}=0,\\
 H_{\a\b a_1\cdots a_5}&=&-2 e^{-2\phi}(\G_{a_1\cdots a_5})_{\a\b},\\
 H_{\a a_1\cdots a_6}&=&  -2 e^{-2\phi}(\G_{a_1\cdots a_6})_\a{}^\b\la_\b,\\
 H_{a_1\cdots a_7}&=& {1\over 3!}\, e^{-2\phi}\ve_{a_1\cdots a_7}{}^{bcd}\left(T_{bcd}-(\G_{bcd})^{\a\b}\la_\a\la_\b+12 S_{bcd}\right).
\ea
Using \eref{1}--\eref{9} one can then show that these forms satisfy the superspace Bianchi--identities/equations of motion \footnote{To prove \eref{h3} and \eref{h7} one must use  the parametrizations \eref{1}--\eref{9}, but  also a set of relations that follow from \eref{1}--\eref{9} through the closure of the susy--algebra of the covariant derivatives $D_a$ and $D_\a$.},
 \ba
  d\w H&=&W,\label{h3}\\
  dH_7&=&0,\label{h7}
 \ea
where $W$ is a form belonging to $V_4$, with $L^{\a\b}$ given simply by,
 \be\label{ls}
L^{\a\b}= S^{\a\b}.
 \ee
Since $W$ is closed, \eref{h3} allows locally to introduce a two--form potential $B$, in which case \eref{h7} is its equation of motion. Viceversa, if one solves \eref{h7} to introduce a six--form potential $B_6$, then \eref{h3} becomes its equation of motion. The ``tilde'' on the three--form curvature $\w H$ will become clear in the next section.

From this construction one concludes the following. To get a consistent solution of ten--dimensional supergravity one must find a super four--form $W$ belonging to $V_4$, expressed in terms of the physical fields; this form determines uniquely a bi--spinor $L^{\a\b}$ that identifies directly the auxiliary field $S^{\a\b}$.  Relevant choices for $W$ are: 1) $W=0$, that gives $S^{\a\b}=0$ and leads to pure supergravity;
2) $W=\a \,{\rm tr FF}$, that belongs to $V_4$ tanks to \eref{ym}, \eref{ym1} and \eref{ym2}, in which case one has,
 \be\label{sym}
S^{\a\b}= -8\,\a \,{\rm tr}(\chi^\a\chi^\b).
 \ee
Indeed, from \eref{ym1}, \eref{ym2}, one obtains $W_{0,4}=W_{1,3}=0$ and,
\be\label{wym}
W_{2,2}=\a\,{\rm tr}(FF)_{2,2}=- 4\,\a \,E^b E^a E^\a E^\b (\G_{[a})_{\a\g}
{\rm tr}(\chi^ \g \chi^\d)(\G_{b]})_{\d\b}.
\ee
Comparing with \eref{v4} one gets then \eref{sym}. This choice for $W$
leads to supergravity coupled minimally to the the super Yang--Mills fields, i.e. the Chaplin--Manton theory. 3) The third solution regards the anomaly canceling theory based on the Bianchi identity \eref{bian}. In this case we must choose, see the next section,
$$
W=\a({\rm tr}FF-K),
$$
for a suitable $K\in V_4$, leading to,
 \be\label{sabc}
S^{\a\b}= -8\,\a \left[{\rm tr}(\chi^\a\chi^\b)- {\rm tr}(T^\a T^\b)\right]+o(\a^2), \quad\quad {\rm tr}(T^\a T^\b)\equiv T_{ab}{}^\a T^{ba\b}.
 \ee
Notice that $S^{\a\b}$ starts at order $\a$. The construction of the four--form $K$ belonging to $V_4$ is most conveniently performed in the framework of the BPT--theorem.

\section{The BPT--theorem}\label{bptth}

The superymmetrization of the Bianchi--identity \eref{bian} does not fit {\it directly} in the scheme of the previous section, since the four--superform ${\rm tr}RR$ does not belong to $V_4$. This is due to the fact, although $d({\rm tr}RR)=0$, for $S^{\a\b}\neq 0$ the gravitational curvature does not satisfy constraints like \eref{ym1} and  \eref{ym2}, and hence $({\rm tr}RR)_{0,4}$ and  $({\rm tr}RR)_{1,3}$ are non vanishing. The BPT--theorem \cite{BPT} overcomes this problem as follows.
\newline
\ul{\it Theorem:}  Given a set of kinematical constraints, like \eref{1}--\eref{9}, there exists an invariant three--superform $X$ and an invariant four--superform $K$ such that,
 \be\label{bpt}
{\rm tr}RR=dX+K, \quad\quad {\rm with}\quad K\in V_4.
 \ee

This theorem guarantees that there exists a consistent solution for ten--dimensional supergravity, if one sets, see \eref{h3},
  \be\label{bian0}
d \w H=\a\,({\rm tr}FF-K)\equiv W.
 \ee
Thanks to the theorem the r.h.s. of this formula belongs, indeed, to $V_4$. Moreover, given \eref{bpt} one can rephrase the relation \eref{bian0} as the desired Bianchi--identity \eref{bian}, now in superspace,
 \be\label{biansuper}
dH=\a\,({\rm tr}FF-{\rm tr}RR), \quad\quad H=\w H-\a X.
 \ee
Thanks to \eref{h03}--\eref{ht} this means that for $H$ one has the constraints,
 \ba
H_{\a\b\g}&=&-\a X_{\a\b\g},\label{hh1}\\
H_{a\a\b}&=& 2(\G_a)_{\a\b} -\a X_{a\a\b}\label{aab}\\
H_{ab\a}&= &-\a X_{ab\a},\label{hh3}
 \ea
while the vectorial torsion is now related to $H$ through,
\be\label{ht1}
T_{abc}=H_{abc}+\a X_{abc}.
\ee
Phrasing \eref{bian0} in this way we can then say, alternatively, that the BPT--theorem provides a consistent set of modified constraints for the $H$--field, in presence of the Lorentz--Chern--Simons form. \eref{biansuper} allows indeed to introduce the super two--form potential $B$ in a standard way as,
 \be\label{b2}
H=dB+\a(\om_{YM}-\om_L),
 \ee
where $\om_{YM}$ and $\om_L$ are respectively the Yang--Mills and Lorentz--Chern--Simons three--forms.

\ul{\it Field redefinitions.} Some comments are in order. First, the proof of the BPT--theorem given in \cite{BPT} uses a set of kinematical constraints that differs heavily from \eref{1}--\eref{9}. However, since all sets of kinematical constraints based on \eref{class} are related by field redefinitions, the proof presented there carries over to the present case.

Next, the expressions of $K$ and $X$ depend clearly on the chosen set of kinematical constraints, but a part from this it is important to realize that even for a fixed chosen set of kinematical constraints the forms $X$ and $K$ are not uniquely determined.
There are indeed two classes of ambiguities, both arising from field redefinitions preserving \eref{1}--\eref{9}, that lead therefore to physically equivalent theories.

I) The first ambiguity amounts to the shift,
 \be\label{shiftb}
X\rightarrow X-dC,
 \ee
where $C$ is an arbitrary two--superform. If $X$ satisfies \eref{bpt} then clearly also $ X-dC$ satisfies this decomposition, with the same $K$. The form $C$ can be absorbed simply by a shift of the two--form potential $B$, $B\rightarrow B+\a\, C$, see \eref{biansuper} and \eref{b2}.

II) The second ambiguity amounts to a shift of the $(1,2)$ component of $X$ of the form,
 \be\label{rescale}
X_{a\a\b}\rightarrow X_{a\a\b}-\La(\G_a)_{\a\b},
 \ee
where $\La$ is an arbitrary scalar superfield. This would lead, instead of \eref{aab}, to,
$$
H_{a\a\b}= 2(1+\a\La/2) (\G_a)_{\a\b} -\a X_{a\a\b},
$$
and the factor $(1+\a\La/2)$ can be eliminated by a rescaling/shift of the super--zehnbeins, accompanied by a shift of the spinorial connection $\Omega_{\a a}{}^b$, see \cite{puzzle}.

\subsection{A realization of the BPT--theorem}

To obtain an explicit superspace formulation for supergravity one must determine the forms $X$ and $K$, on the basis of \eref{bpt}.

As shown in \cite{BPT} it is sufficient to realize the decomposition ${\rm tr} RR=dX+K$ in the sectors $(0,4)$ and $(1,3)$, as it will then hold automatically also in the sectors $(2,2)$, $(3,1)$ and $(4,0)$. Notice in particular that once one has found a three--form $X$ such that $K$ vanishes in the sectors (0,4) and (1,3), the form $K$ is automatically closed because ${\rm tr} RR$ is a closed form. In summary, to find an explicit realization of \eref{bpt} it is  sufficient to solve the equations,
 \ba
 ({\rm tr}RR)_{0,4}&=&(dX)_{0,4},\label{rr1}\\
({\rm tr}RR)_{1,3}&=& (dX)_{1,3}.\label{rr2}
 \ea
The algebraic details needed for the solutions of these equation are given in the appendix, here we repeat only the main steps.

We start with equation \eref{rr1}. It involves on its r.h.s. only the components $X_{0,3}$ and  $X_{1,2}$, and on its l.h.s only the curvature components \eref{5}. This equation admits non trivial solutions for a vanishing  $X_{0,3}$, i.e. for \footnote{Here we are interested only in "minimal" solutions, i.e. solutions that correspond just to the supersymmetric completion of the Bianchi--identity \eref{bian}; these solutions do not include, for example, the terms quartic in the curvature with an irrational coefficient, of the form $\zeta(3)R^4$, that are present in the low energy effective action of all ten dimensional supergravity theories \cite{grzan}. As long as one insists on \eref{class}, to supersymmetrize such terms one must indeed choose a non vanishing $X_{0,3}$, see \cite{r4}.},
 \be \label{03}
X_{\a\b\g}=0.
 \ee
With this choice the general solution of \eref{rr1} for $X_{1,2}$ is given by,
 \be\label{xgen}
X_{a\a\b}= -72 \G^b_{\a\b}S^2_{ab}+\La(\G_a)_{\a\b}+(\G^b)_{\a\b}C_{ab}+(\G_{abcde})_{\a\b}Y^{bcde},
 \ee
where we set $S^2_{ab}\equiv S_{acd}S_b{}^{cd}$, while the scalar field $\La$ as well as the antisymmetric tensors $C_{ab}$ and $Y_{abcd}$ are completely arbitrary. The fields $C_{ab}$ and $\La$ can be eliminated respectively through the shifts \eref{shiftb} and \eref{rescale}, so that we can take as general solution of \eref{rr1},
 \be\label{x12}
X_{a\a\b}= -72 \,\G^b_{\a\b}S^2_{ab}+(\G_{abcde})_{\a\b}Y^{bcde},
 \ee
where the tensor $Y^{bcde}$ is still arbitrary. The  simplest solution would amount to set,
$$
Y^{bcde}=0,
$$
but the key point is that for such a choice the equation \eref{rr2} would not admit any solution for $X_{2,1}$ at all! Stated differently, it is not sufficient to solve the $H$--Bianchi identity \eref{biansuper} at the level $(0,4)$ -- at least when one takes corrections beyond the first order in $\a$ into account. This point remains still controversial w.r.t. to the perturbative approach \cite{gates2}.
The equation \eref{rr2} involves actually $X_{1,2}$ as well as $X_{2,1}$, and it can be seen that it admits a solution for $X_{2,1}$ if one chooses, see the appendix,
 \be\label{y4}
Y_{abcd}=2D_{[a}S_{bcd]}+4S_{[abc}D_{d]}\phi+6(ST)_{[abcd]},
 \ee
where $(ST)_{abcd}\equiv S_{eab}T_{cd}{}^e$. Once a solution for $X_{2,1}$ of eq. \eref{rr2} exists, it can also be shown to be {\it unique}, and once \eref{rr1} and \eref{rr2} are solved, also $X_{3,0}$ and the form $K\in V_4$ are {\it consistently and uniquely determined}, see \cite{BPT}.

In summary we can say that the decomposition ${\rm tr}RR=dX+K$ fixes uniquely the superforms $X$ and $K$, once one has chosen \eref{03} and \eref{x12} with \eref{y4}.
At this point it is a straightforward, but very lengthy, exercise to determine $X_{2,1}$, $X_{3,0}$ and $K$ explicitly. In particular we know that $K_{2,2}$ has the form,
\be\label{k22}
K_{2,2}={1\over 2}\,E^bE^aE^\a E^\b (\G_{[a})_{\a\g}K^{\g\d}(\G_{b]})_{\d\b},
\ee
and the relation $W=\a\,({\rm tr}FF-K$) at level (2,2) gives then,
 \be\label{sab}
S^{\a\b} =-8\,\a \,{\rm tr}(\chi^\a\chi^\b)- \a K^{\a\b},
 \ee
that determines the auxiliary field $S^{\a\b}$. Similarly the equation \eref{ht1} determines the auxiliary field $T_{abc}$.
Actually, the bi--spinor $K^{\a\b}$ as well as the field $X_{abc}$ are complicated functions of the physical fields, as well as of the auxiliary fields $S^{\a\b}$ and $ T^{abc}$ themselves. This means that the equations for those fields become implicit relations that are better written as,
 \ba
S^{\a\b} &=&-8\,\a \,{\rm tr}(\chi^\a\chi^\b)- \a \,K^{\a\b}(S,T),\\
T_{abc}&=&H_{abc}+\a\, X_{abc}(S,T),
 \ea
and hence they can only be solved iteratively order by order in $\a$.

We insist, however, on the fact that our construction leads to {\it a theory that is supersymmetric to all orders in $\a$}, due to the algebraic consistency of our parametrizations in all sectors of the Bianchi--identities.

\subsection{The theory at first order in $\a$}

At first order in $\a$ our construction becomes particularly simple. Since the field $S^{\a\b}$ is of order $\a$, from \eref{x12} and \eref{y4} we see that also $X_{1,2}$
is of order $\a$. Since $({\rm tr} RR)_{1,3}$ in \eref{rr2} is also of order $\a$ -- see \eref{5} and \eref{6} -- this equation implies then that also $X_{2,1}$ is of order $\a$. To analyze the orders of $X_{3,0}$ and $K$ we write the decomposition \eref{bpt} at level (2,2), see \eref{k22},
 \be\label{22}
({\rm tr}RR)_{2,2}=(dX)_{2,2}+{1\over 2}\,E^bE^aE^\a E^\b (\G_{[a})_{\a\g}K^{\g\d}(\G_{b]})_{\d\b}.
 \ee
At level (2,2) ${\rm tr}RR$ has now also a term at zero order in $\a$, in that \eref{5} and \eref{6} give,
$$
{\rm tr}(RR)_{2,2}=- 4\,E^b E^a E^\a E^\b (\G_{[a})_{\a\g}
{\rm tr}(T^ \g T^\d)(\G_{b]})_{\d\b}+o(\a).
$$
On the other hand, since all components of $X$, a part from possibly $X_{3,0}$, are of order $\a$ we have,
$$
(dX)_{2,2}={1\over 2}\,E^aE^bE^\a E^\b \G^c_{\a\b}X_{cba}+o(\a).
$$
Substituting these expressions in \eref{22} we conclude that also $X_{3,0}$ is of order $\a$, and that we have,
$$
K^{\a\b}=-8\,{\rm tr}(T^ \a T^\b) +o(\a).
$$
Substituting this expression in \eref{sab} gives \eref{sabc}.

Since the form $X$ is entirely of order $\a$, the corrections to the $H$--constraints   \eref{hh1}--\eref{hh3} all vanish at this order and the superspace parametrizations become extremely simple. We collect here the most important ones,
 \ba
H_{\a\b\g}&=&0,\label{abg}\\
H_{a\a\b}&=& 2(\G_a)_{\a\b} +o(\a^2)\label{aab2} \\
H_{ab\a}&= &o(\a^2),\\
T_{abc}&=&H_{abc}+o(\a^2),\\
S^{\a\b}&=& -8\,\a \left({\rm tr}(\chi^\a\chi^\b)- {\rm tr}(T^\a T^\b)\right)+o(\a^2),\label{dt1}\\
S_{ab}{}^\a&=&-4\,\a \left({\rm tr}(F_{ab}\chi^\a)- {\rm tr}(R_{ab} T^\a)\right)+o(\a^2),\label{dt3}
 \ea
where ${\rm tr}(R_{ab} T^\a)\equiv R_{ab}{}^{cd}T_{dc}{}^\a$.  \eref{dt3} follows from \eref{dt1} through the  defining relation \eref{9}, using \eref{ym3} and its gravitational analog,
$$
D_\a T_{ab}{}^\b=T_{\a\ve}{}^\b T_{ab}{}^\ve+{1\over 4}\,(\G^{cd})_\a{}^\b R_{cdab}+o(\a),
$$
that is a direct consequence of the torsion Bianchi--identity \eref{tbian}.
Notice the symmetry between the supergravity and the Yang--Mills first order corrections in \eref{dt1} and \eref{dt3}.

\section{Relation with previous superspace formulations}

The first realization of the BPT--theorem has been given in \cite{BPT}, in the framework of a particular set of kinematical constraints. The set used in that reference was very convenient for the proof of the theorem, but it was not particularly suitable for
a perturbative expansion in $\a$, due to its cumbersome ``mixing'' between physical and auxiliary fields.

A more convenient set of constraints was proposed in reference \cite{raciti} --  relying  again on the non--perturbative approach and on the BPT--theorem -- with the particular aim of comparing the perturbative approach of \cite{gates2} with the non--perturbative one, revealing a disagreement already at first order in $\a$, as mentioned in the introduction.
The constraints used in \cite{raciti} differ from \eref{1}--\eref{9} by a shift of the vectorial connection, $\Omega_{ab}{}^c\ra \Omega_{ab}{}^c-T_{ab}{}^c$, and by some trivial rescalings. After these simple transformations the three--form $X$ found by the authors of \cite{raciti}  -- in the following called $X^*$ -- becomes again (necessarily) of the form \eref{03}, \eref{x12},
 \ba
 X^*_{\a\b\g}&=&0,\\
X^*_{a\a\b}&=& -72 \,\G^b_{\a\b}S^2_{ab}+(\G_{abcde})_{\a\b}Y^{*bcde}. \label{x12z}
 \ea
However, the choice made for $Y^*_{abcd}$ in \cite{raciti} differs from \eref{y4} in that \footnote{This expression is obtained from formula (24) of ref. \cite{raciti} by setting $\Omega^*_{ab}{}^c= \Omega_{ab}{}^c-T_{ab}{}^c$, and using the identity $R_{[abcd]}=D_{[a}T_{bcd]}+T^2_{[abcd]}$, that follows from \eref{tbian}.  A part from this one has to take into account also the different overall normalizations of $S^{\a\b}$ and $H$ used in \cite{raciti}, w.r.t. the present paper.},
 \ba
 Y^*_{abcd}&=&{1\over 3}\,\left(D_{[a}T_{bcd]}+2T^2_{[abcd]}-T_{[ab}{}^\a (\G_{cd]})_\a{}^\b\la_\b\right) + 4S_{[abc}D_{d]}\phi \nn\\
 &&+ (ST)_{[abcd]} +6S^2_{[abcd]}
+{1\over 72}\,\ve_{abcdc_1\cdots c_6}\,S^{c_1c_2c_3}T^{c_4c_5c_6},\label{y4z}
 \ea
where $S^2_{abcd}=S_{abe}S^e{}_{cd}$. It can indeed be shown that with the choice \eref{y4z} the equation \eref{rr2} allows a consistent solution for $X^*_{2,1}$, and hence the forms $X^*$ and $K^*$ are uniquely and consistently determined, as in the previous section, by the relation,
 \be\label{bptz}
{\rm tr}RR=dX^*+K^*, \quad\quad K^*\in V_4.
 \ee

With respect to \eref{y4} the expression \eref{y4z} appears more complicated, but the most important difference is that  $Y^*_{abcd}$ and $X^*_{a\a\b}$ are {\it of order zero in $\a$}, while \eref{x12} is of first order in $\a$. This means that with the choice \eref{y4z} the constraint for $H_{a\a\b}$ in \eref{aab} has now necessarily a non--vanishing first order correction in $\a$, as opposed to \eref{aab2}. This feature was on the basis of the disagreement between \cite{raciti} and the perturbative approach
\cite{gates2}, in that the letter claimed for a {\it vanishing first order correction} to this constraint, which -- in turn -- is now in agreement with \eref{aab2}. We will now show that the solution of \cite{raciti} is related to the one presented in this paper by a (rather complicated) field--redefinition, and that the two solutions are thus physically equivalent. This reconciles in particular the first order results of the perturbative approach \cite{gates1}--\cite{gates3}, with the ones of the non--perturbative one.

\subsection{Equivalence between two non--perturbative solutions}

The proof of the physical equivalence of the decompositions \eref{bpt} and \eref{bptz} relies on the field redefinitions \eref{shiftb} and \eref{rescale} introduced in section \ref{bptth}. Since for a vanishing $X_{0,3}$ the forms $X$ and $K$ are uniquely fixed by the component $X_{1,2}$, it is sufficient to find a two--form $C$ and scalar field $\La$ such that the expressions \eref{x12} and \eref{x12z} transform into each other, keeping $X_{0,3}$ vanishing.

To this order we define a super two--form $C$ with components,
 \ba
 C_{\a\b}&=&0,\label{c1}\\
 C_{a\a}&=& -{2\over 3}\,(\G_a\G^{bc})_{\a\b}T_{bc}{}^\b,\label{c2}\\
 C_{ab}&=&-{4\over 3}\,R_{[ab]}-{1\over 3}\,T_{cd}{}^\a (\G_{ab}{}^{cd})_\a{}^\b\la_\b-
       10\,D^cS_{abc}-22\, S_{cd[a}T_{b]}{}^{cd},\label{c3}
 \ea
where $R_{[ab]}$ is the antisymmetric part of the Ricci tensor $R_{ab}=R^c{}_{acb}$. Notice that this two--form is of zero order in $\a$, as is $X^*_{1,2}$. Using \eref{1}--\eref{9} it is then a lengthy but straightforward calculation to show that one has,
 \ba
  (dC)_{\a\b\g}&=&0,\\
  (dC)_{a\a\b}&=&(\G_a{}^{bcde})_{\a\b}\left[{1\over 3}\left(D_{b}T_{cde}+2T^2_{bcde}-T_{bc}{}^\a (\G_{de})_\a{}^\b\la_\b\right)-5\,(ST)_{bcde}\right.\nn\\
&&+6\,S^2_{bcde}-2D_b\,S_{cde}
\left.+{1\over 72}\,\ve_{bcdec_1\cdots c_6}\,S^{c_1c_2c_3}T^{c_4c_5c_6}\right] +(\G_a)_{\a\b}\La,\label{dc12}
 \ea
where we defined the superscalar,
  \be\label{ss}
 \La={2\over 3}\,R-36 S_{abc}S^{abc}+2 S_{abc}T^{abc}, \quad\quad R\equiv R_a{}^a.
  \ee
Subtracting \eref{dc12} from \eref{x12z} one obtains then,
\ba
 (X^*-dC)_{\a\b\g}&=& 0,\\
 (X^*-dC)_{a\a\b}-(\G_a)_{\a\b}\La &=& X_{a\a\b},
\ea
with $X_{a\a\b}$ given in \eref{x12}. From this we conclude that the solution proposed in \cite{raciti} is physically equivalent to ours.

\section{Concluding remarks}

In this paper we have proposed a new all order solution of the superspace Bianchi--identities for (minimal) anomaly free $N=1$, $D=10$ supergravity. The new solution realizes the physical equivalence -- at first order -- between the perturbative approach of \cite{gates1}--\cite{gates3}, and the non--perturbative approach of \cite{Torino}--\cite{11}. Eventually the set \eref{1}--\eref{9}, with \eref{abg}--\eref{dt3}, represents a further simplification of the first order constraints of \cite{gates2}, achieved by a suitable torsion shift.

Although we performed only an explicit first order calculation, given \eref{x12} with the (closed) expression \eref{y4} the derivation of the higher order theory is now a merely technical point. Notice in particular that the BPT--theorem entails also closed expressions for $X_{2,1}$, $X_{3,0}$ and $K$, in terms of the auxiliary fields $S^{\a\b}$ and $T^{abc}$ and their spinorial derivatives.

Since at zero order in $\a$ the Yang--Mills and supergravity supercurvatures have identical parametrizations, see \eref{fr},
the ``difference--structure'' of the term ${\rm tr} FF-{\rm tr}RR$ in the $H$--Bianchi identity, entails the difference--structure of the first order corrections in \eref{dt1} and \eref{dt3}. Since exactly this feature was the starting point of the component level Noether--procedure of \cite{Bergs}, our new superspace formulation is suitable for a direct comparison with the higher order results proposed in that reference.

\ul{\it Poltergeists and Gauss--Bonnet action.} The choice \eref{x12z} of ref. \cite{raciti} is characterized by a further problematic feature, i.e. the presence of a term {\it linear} in the derivative of the curvature of the physical field $T_{abc}=H_{abc}+o(\a)$,
 \be\label{lin}
(X^*_{a\a\b})_{\rm lin}={1\over3}\,(\G_{abcde})_{\a\b}D^bT^{cde}.
 \ee
It can be seen that this terms would lead in $X^*_{3,0}$ to a zero order contribution that is linear in $ \Box T_{abc}$, and hence \eref{ht1} would have the structure,
$$
(1+c\a \Box)T_{abc}=H_{abc}+ \cdots,
$$
where $c$ is a numerical constant, and the dots indicate non linear terms. Due to supersymmetry this would then lead to an Einstein equation of the form,
$$
(1+c\a\Box)R_{ab}=j_{ab},
$$
where $j_{ab}$ is a non linear current. The presence of the fourth--order derivatives on the zehnbein would then imply the propagation of unphysical poltergeist degrees of freedom in the metric, and similar poltergeists would appear also in the fermionic fields. The same problematic feature occurred also in the original version of the non--perturbative approach \cite{BPT}.
In the present version of the theory these unphysical modes are absent, because in \eref{x12} there are no terms that are linear in the physical fields. The disappearance of these linear terms is easily understood from the point of view of the field redefinitions \eref{c2}, \eref{c3}.  Indeed, \eref{c2} is linear in the field strength of the gravitino, and \eref{c3} -- that shifts $B_{ab}$ -- contains a term that is linear in the field strength of $B_{ab}$ in that,
 $$
R_{[ab]}=-{1\over 2}D^cT_{abc}=-{1\over 2}D^cH_{abc}+o(\a).
$$
Similarly the scaling parameter \eref{ss} contains a term that is linear in the second derivative of the metric. These linear transformations eliminate thus, in particular, the linear term \eref{lin} \footnote{Strictly speaking these transformations -- being linear in the {\it derivatives} of the fields -- are non--invertible, and hence they are only algebraically allowed.}. On the other hand in \eref{x12} there appears a term that is linear in the derivative of the auxiliary field $S_{abc}$. However, since this field is  at first order in $\a$ already quadratic in the physical fields, see \eref{sabc}, its iterative determination will never give rise to terms that are linear in the physical fields.  In the formulation of the theory presented in this paper all equations of motion are therefore free from poltergeists.

This holds in particular also for Einstein's equation, that at first order in $\a$  gets then corrections that are quadratic in the Riemann tensor \footnote{Since in the action there is a term quadratic in $F_{ab}$ -- the Yang--Mills action -- there are necessarily also terms quadratic in $R_{abcd}$. This is due to the difference--structure of \eref{dt1} and \eref{dt3}.}, without fourth--order derivative terms like $\Box R_{ab}$. This implies  that the corresponding term in the action is necessarily the Gauss--Bonnet action, that in the language of differential forms can be written as the integral of a ten--form,
 $$
 S_{\rm GB}=\a\int \ve_{a_1\cdots a_{10}}\,E^{a_1}\cdots E^{a_6}\,R^{a_7a_8}\,R^{a_9a_{10}}.
 $$
This form is indeed naturally predicted by the low energy effective actions provided by string theory calculations \cite{grzan}. Given the simplicity of our superspace construction we hope to be able to derive the complete supersymmetrization of $S_{\rm GB}$ in ten dimensions to order $\a$, that is still unknown.

As we have observed several times, the construction given in this paper furnishes an all order solution of the superspace Bianchi--identity \eref{bian}, that is based on the classical torsion constraint \eref{class}. On the other hand, the order--$\a^2$ solution proposed recently in \cite{reilly} relaxes this constraint and allows for a torsion of the form, in our notations,
 \be \label{1050}
T_{\a\b}{}^a=2\G^a_{\a\b}+k\,\a\,(\G_{bcdef})_{\a\b}T^{abc}S^{def},
 \ee
where
$k$ is a fixed numerical coefficient. A direct comparison between our construction and the one of \cite{reilly} is rather difficult, due to the complicated formulas that follow from \eref{1050}. A priori it could happen that the two approaches are related by field redefinitions, but for this it would be necessary that the second term in \eref{1050} can be eliminated by a field redefinition, and by inspection this is rater unlikely. The two solutions seem thus unrelated. A drawback of the solution in \cite{reilly} is that, once one renounces to \eref{class}, it is almost impossible to keep
the algebraic consistency of the entire construction under control. Here we do not want to move any concrete criticism against a solution based on \eref{1050} in that -- apart from consistency requirements -- eventually only an explicit superspace string calculation can decide which formulation is physically correct. Our point is simply that, from an algebraic point of view, there is no {\it  need} to modify the classical torsion constraint, since there exists a well defined solution based on \eref{class}: the ``minimal solution'' presented in this paper.

To conclude we observe that, as shown in  \cite{r4}, there are ``non--minimal'' corrections to the string effective action of higher order in $\a'$ -- like the term $\zeta(3)R^4$ -- that can likewise be accommodated  in the present framework in terms of an appropriate four--form $W$ (see \eref{v4} and \eref{ls}), preserving thus once more the classical rigid torsion constraint \eref{class}.

\vskip1truecm

\paragraph{Acknowledgements.}
The authors would like to thank warmly Osvaldo Chand\'ia for his participation at the earlier stage of this work. This work is supported in part by the European Community's Human Potential Programme under contract MRTN-CT-2004-005104,
``Constituents, Fundamental Forces and Symmetries of the Universe",
and by the INTAS Project Grant 05-1000008-7928.

\vskip1truecm

\section{Appendix: The decomposition ${\rm tr} RR=dX+K$}

As mentioned in the text, to prove the decomposition in the title it is sufficient to solve \eref{rr1} and \eref{rr2}.

\ul{\it Solution of \eref{rr1}.}\quad  Since $X_{0,3}=0$, equation \eref{rr1} amounts to,
\be
R_{(\a\b cd}\,R_{\g\d)}{}^{dc}=-144\, S^2_{ab}\,\G^a_{(\a\b}\,\G^b_{\g\d)}=2\,\G^a_{(\a\b}X_{a\g\d)},
\ee
where on the l.h.s. we inserted \eref{5}, and the symmetrization is intended over the indices $(\a\b\g\d)$. The general solution of this equation has the structure,
 \be\label{a1}
X_{a\a\b}= -72 \,\G^b_{\a\b}S^2_{ab}+  (\G_a)_{\g(\a}X^\g{}_{\b)},
 \ee
where $X^\g{}_\b$ is an arbitrary by--spinor. This is due to the cyclic identity,
 \be\label{cyclic}
(\G_a)_{(\a\b}(\G^a)_{\g)\d}=0.
 \ee
Since one has the general representation,
$$
X^\g{}_\b=\d^\g{}_\b\La+{1\over 2}\,(\G_{ab})^\g{}_\b C^{ab} +(\G_{abcd})^\g{}_\b Y^{abcd},
$$
\eref{a1} reduces to \eref{xgen}.

\ul{\it Solution of \eref{rr2}.} \quad Equation \eref{rr2} can be written as,
 \be\label{a2}
R_{a(\a cd}\,R_{\b\g)}{}^{dc}+{1\over 2}\, D_{(\a}X_{a\b\g)}+{1\over 2}\,T_{(\a\b}{}^\d X_{a\g)\d}=\G^b_{(\a\b}X_{ba\g)},
 \ee
where the unknown is the field $X_{ba\g}$, antisymmetric in $a$ and $b$, and the symmetrization is over $(\a\b\g)$. On general grounds the l.h.s. of this equation has the structure,
 \be\label{decom}
\G^b_{(\a\b}W_{ba\g)} +(\G_a{}^{c_1c_2c_3c_4})_{(\a\b}W_{\g)c_1c_2c_3c_4},
 \ee
i.e. contains terms that factorize a  $\G_1$--matrix and terms that factorize a $\G_5$--matrix. The equation \eref{a2} can have a solution for $X_{ba\g}$ only if the  terms that factorize a $\G_5$ drop eventually out, in which case one has the (unique) solution $X_{ba\g}=W_{ba\g}$. We insert thus \eref{xgen} in \eref{a2} and keep only the terms that factorize a $\G_5$. An explicit calculation gives,
 \be\label{gamma5}
\left[ R_{a(\a cd}\,R_{\b\g)}{}^{dc}+{1\over 2}\, D_{(\a}X_{a\b\g)}+{1\over 2}\,T_{(\a\b}{}^\d X_{a\g)\d}\right]_{\G_5}=(\G_a{}^{c_1c_2c_3c_4})_{(\a\b}W_{\g)c_1c_2c_3c_4},
 \ee
where,
 \be
W_{\g c_1c_2c_3c_4}\equiv
\G^b_{\g\d}\left(4\,{\w T}_{b[c_1}{}^\d \,S_{c_2c_3c_4]}+6\,{\w T}_{[c_1c_2}{}^\d \, S_{c_3c_4]b}\right) +\left({1\over 2}\,D_{\g}+\la_{\g}\right)Y_{c_1c_2c_3c_4}, \label{1440}
 \ee
and we defined,
$$
{\w T}_{ab}{}^\a=T_{ab}{}^\a  -{1\over 4}\, S_{ab}{}^\a.
$$
The question is now whether there exists an antisymmetric tensor $Y_{abcd}$ such that
the expression in \eref{gamma5} factorizes eventually a $\G_1$--matrix. For this to happen it is necessary and sufficient that  \eref{1440} assumes the form,
 \be\label{no1440}
W_{\g c_1c_2c_3c_4}=(\G_{[c_1})_{\g\d}Z^\d_{c_2c_3c_4]},
 \ee
for some antisymmetric tensor $Z^\d_{c_2c_3c_4}$ \footnote{From a group theoretical point this is equivalent to require that there exists a
tensor $Y_{abcd}$ such that from $W_{\g c_1c_2c_3c_4}$ all irreducible representations of dimension 1440 of $SO(1,9)$ drop out. These representations correspond to spinorial tensors of the form $W_{\g c_1c_2c_3c_4}$, completely antisymmetric in the four vectorial indices, with all $\G$--traces vanishing.}. Indeed, thanks to the cyclic identity \eref{cyclic} in this case one has,
$$
(\G_a{}^{c_1c_2c_3c_4})_{(\a\b}W_{\g)c_1c_2c_3c_4}=-{1\over 2}\G^b_{(\a\b}(\G^{c_1c_2c_3}{}_a\G_b)_{\g)\d} Z^\d_{c_1c_2c_3},
$$
that is of the form of the first term in \eref{decom}.

An explicit evaluation of $\left({1\over 2}\,D_{\g}+\la_{\g}\right)Y_{c_1c_2c_3c_4}$ shows that the reduction \eref{no1440} happens indeed if one chooses for $Y_{c_1c_2c_3c_4}$  \eref{y4} or \eref{y4z}. The conclusion is that for both these choices
the equation \eref{a2} admits a consistent and unique solution.

\end{document}